\documentclass[usenatbib]{mn2e}
\usepackage{epstopdf}
\usepackage{graphicx}
\usepackage{subfig}
\usepackage{color}

\title[GRB universal scaling]{Update on the GRB universal scaling E$_{\rm{X,iso}}$-E$_{\rm{\gamma,iso}}$-E$_{\rm{pk}}$ with ten years of \textit{Swift} data.}

\author[E. Zaninoni, M. G. Bernardini, R. Margutti, and L. Amati]{E. Zaninoni$^{1}$\thanks{E-mail: elena.zaninoni@gmail.com (EZ)}, M. G. Bernardini$^{2}$, R. Margutti$^{3}$, and L. Amati$^{4}$\\
$^{1}$ ICRANet-Rio, Centro Brasileiro de Pesquisas F\'isicas, Rua Dr. Xavier Sigaud 150, 22290--180 Rio de Janeiro, Brazil\\
$^{2}$ INAF - Osservatorio Astronomico di Brera, via Bianchi 46, Merate 23807, Italy\\
$^{3}$ Harvard-Smithsonian Center for Astrophysics, 60 Garden Street, Cambridge, MA02138\\
$^{4}$ INAF, Istituto di Astrofisica Spaziale e Fisica Cosmica, Bologna, Via Gobetti 101, I-40129 Bologna, Italy}

\begin{document}

\date{Accepted . Received ; in original form }

\pagerange{\pageref{firstpage}--\pageref{lastpage}} \pubyear{2015}

\maketitle

\label{firstpage}

\begin{abstract}
From a comprehensive statistical analysis of \textit{Swift} X-ray light-curves of gamma-ray bursts (GRBs) collected from December 2004 to the end of 2010, we found a three-parameter correlation between the isotropic energy emitted in the rest frame 1-10$^4$ keV energy band during the prompt emission (E$_{\rm{\gamma,iso}}$), the rest frame peak of the prompt emission energy spectrum (E$_{\rm{pk}}$), and the X-ray energy emitted in the rest frame 0.3-30 keV observed energy band (E$_{\rm{X,iso}}$), computed excluding the contribution of the flares. In this paper, we update this correlation with the data collected until June 2014, expanding the sample size with $\sim$35\% more objects, where the number of short GRBs doubled. With this larger sample we confirm the existence of a universal correlation that connects the prompt and afterglow properties of long and short GRBs. We show that this correlation does not depend on the X-ray light-curve morphology and that further analysis is necessary to firmly exclude possible biases derived by redshift measurements. In addition we discuss about the behavior of the peculiar objects as ultra-long GRBs and we propose the existence of an intermediate group between long and short GRBs. Interestingly, two GRBs with uncertain classification fall into this category. Finally, we discuss the physics underlying this correlation, in the contest of the efficiency of conversion of the prompt $\gamma$-ray emission energy into the kinetic energy of the afterglow, the photosferic model, and the cannonball model.
\end{abstract}



\section{Introduction}
The \textit{Swift} satellite \citep{2004ApJ...611.1005G},  launched on November 2004, opened a new era for the study and understanding of gamma-ray busts (GRBs), detecting more than 900 objects until January 2015. Thanks to its unique observing capabilities,  many correlations involving prompt and afterglow emission quantities could be further investigated (e.g. \citealt{2001ApJ...552...57R,2002A&A...390...81A,2004ApJ...616..331G,2004ApJ...609..935Y,2008MNRAS.391L..79D}). 

One of the most studied is the Amati relation \citep{2002A&A...390...81A} that involves the isotropic energy emitted in the rest frame 1-10$^4$  keV energy band during the prompt emission (E$_{\rm{\gamma,iso}}$) and the photon energy at which the prompt emission energy spectrum peaks (E$_{\rm{pk}}$). This relation is followed by long GRBs, while short GRBs lie in a separate region of the E$_{\rm{\gamma,iso}}$-E$_{\rm{pk}}$ plane. Recent papers (e.g. \citealt{2012ApJ...755...55Z,2014arXiv1412.5630S,2015MNRAS.448..403C}) showed that short GRBs also follow a well defined relation in the E$_{\rm{\gamma,iso}}$-E$_{\rm{pk}}$ plane, with similar slope but different normalization with respect to long GRBs.

From a comprehensive statistical analysis of \textit{Swift} X-ray light-curves collected from December 2004 until December 2010 (\citealt{2013MNRAS.428..729M}, hereafter M13), we found a three-parameter correlation  between E$_{\rm{\gamma,iso}}$, E$_{\rm{pk}}$, and the X-ray energy emitted in the rest frame 0.3-30 keV observed energy excluding the flare activity (E$_{\rm{X,iso}}$). The uniqueness of this correlation is that it accommodates long, short, and low-energy GRBs in a single scaling, involving prompt and afterglow emission quantities (\citealt{2012MNRAS.425.1199B}, hereafter B12, M13). This finding suggests that its physical origin is deeply connected with properties that are shared by the GRBs as a whole.

In this paper we update the E$_{\rm{X,iso}}$-E$_{\rm{\gamma,iso}}$-E$_{\rm{pk}}$ correlation including all GRBs observed by \textit{Swift} until June 2014. We select the sample following the same prescriptions as in B12 and M13. In particular, we consider only GRBs with: i) secure redshift measurement; ii) measured E$_{\rm{pk}}$; iii) complete X-ray light-curve. The new sample contains about 35\% more GRBs than the old one, and, in particular, twice the number of short GRBs. This new sample gives us the possibility to better investigate: i) the role of short GRBs in the E$_{\rm{X,iso}}$-E$_{\rm{\gamma,iso}}$-E$_{\rm{pk}}$ correlation; ii) the possible link between long and short GRBs; iii) the properties of the group of GRBs that lies between long and short GRBs (which we call \textit{intermediate group}); iv) the characteristics of peculiar GRBs, like ultra-long GRBs; v) the relation between prompt and afterglow emission.

This paper is organized as follow: description of criteria used for the sample selection (Sec. \ref{sec:sample}); description of the E$_{\rm{X,iso}}$-E$_{\rm{\gamma,iso}}$-E$_{\rm{pk}}$ correlation and of particular GRBs included in the sample (i.e. short GRBs, ultra-long GRBs and GRBs with uncertain classification); Sec. \ref{sec:three}); discussion of intermediate group, possible biases, and the physical motivations for this correlation (Sec. \ref{sec:discussion}); summary and conclusions (Sec. \ref{sec:summary}). Uncertainties are given at 68\% confidence level (c.l.) unless explicitly mentioned. Standard cosmological quantities have been adopted: H$_0$ = 70 km s$^{-1}$ Mpc$^{-1}$, $\Omega_{\rm{\Lambda}}= 0.7$ and $\Omega_{\rm{M}}=0.3$.


\section{Sample selection}
\label{sec:sample}
We updated our sample (B12, M13) selecting all GRBs observed until June 2014 that fulfill the following requirements (M13):
\begin{enumerate} 
\item Their reshifts $z$ are derived from optical spectroscopy or they have photometric redshifts for which potential sources of degeneracy (e.g. dust extinction) can be ruled out with high confidence.
\item  It was possible to measure the rest-frame peak energy (E$_{\rm{pk}}$) and prompt emission isotropic energy in the rest-frame 1-10$^4$ keV energy band (E$_{\rm{\gamma,iso}}$ \citealt{2006MNRAS.372..233A,2008MNRAS.391..577A,2009A&A...508..173A}) from the broadband modeling of the prompt emission time-integrated spectrum.
\item They were observed by \textit{Swift}/XRT and have a complete X-ray light-curve, i.e. promptly re-pointed by \textit{Swift}/XRT ($t_{\rm{rep}}<300$ s) and for which observations were not limited by any observing constraint.
\end{enumerate}
We obtained a new sample composed of 81 long GRBs, 11  short GRBs and 2 GRBs with uncertain classification\footnote{The GRBs with uncertain classification are GRB 090426 and GRB 100816A. A detailed discussion can be found in Sec. \ref{sec:threethree}.}. The new sample contains $\sim$35\% more GRBs than the previous one; in particular the sample of short GRBs doubled (Table \ref{tab:sample}).  

\begin{table}
\caption{List of 33 GRBs added to the old sample. Short GRBs are marked in boldface, while the GRB with uncertain classification is underlined.}
\label{tab:sample}
\begin{tabular}{p{\columnwidth}}
\hline
GRB name\\
\hline
\textbf{080123}, \underline{090426},  \textbf{100117A}, \textbf{100625A}, 110106B, 110205A, 110213A, 110503A,
 110715A, 110731A, 110801A, 110818A, 111107A, \textbf{111117A}, 111209A, 111228A, 120119A, 120326A, 120712A, 120802A, 120811C, 121128A, 130408A, 130427A, 130505A, \textbf{130603B}, 130701A, 
130831A, 130907A, 130925A, 131030A, 140206A, 140419A \\ 
\hline
\end{tabular}
\end{table}

\begin{figure*}
\centering
\includegraphics[width=\textwidth]{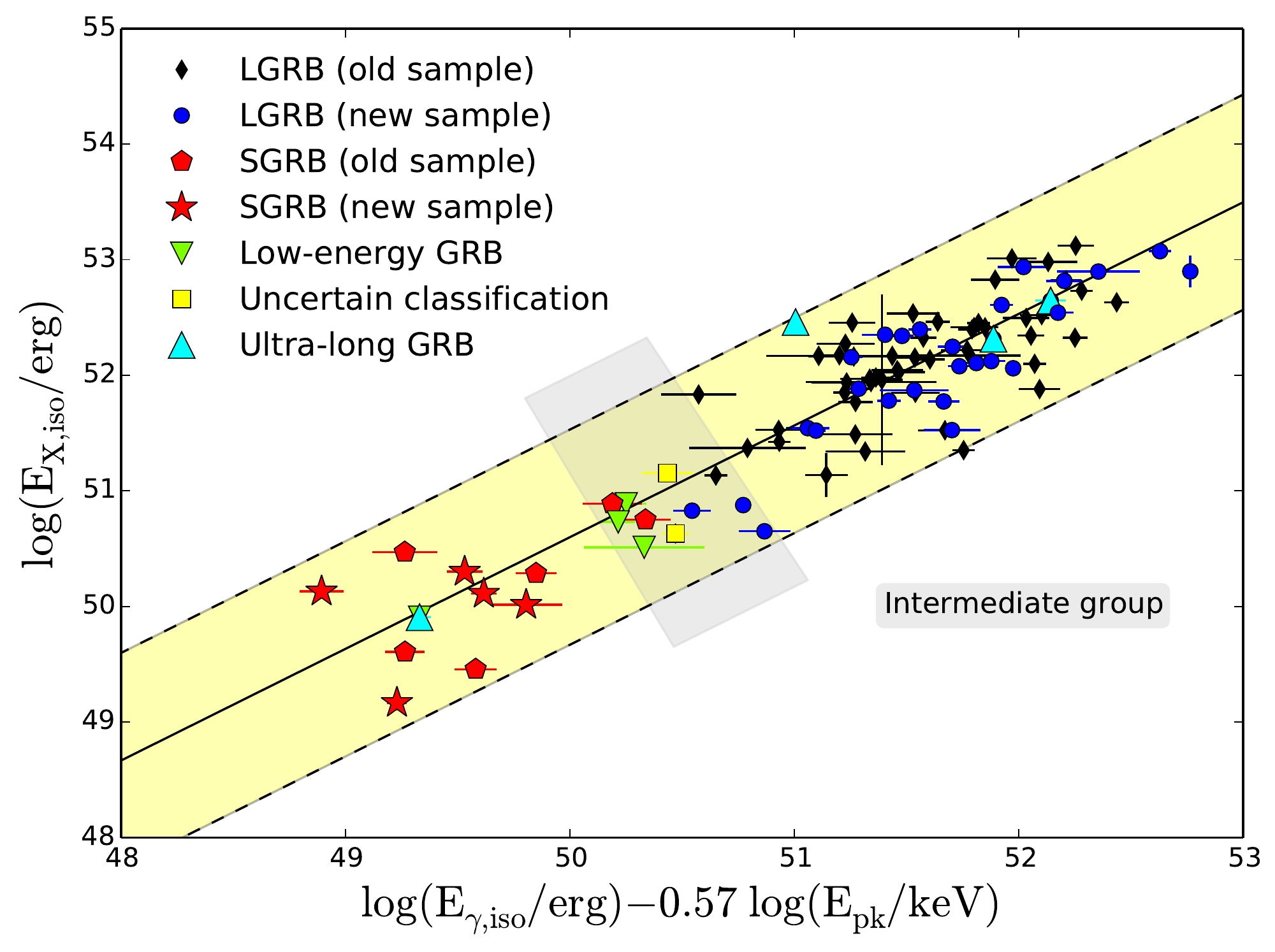}
\caption{E$_{\rm{X,iso}}$ - E$_{\rm{\gamma,iso}}$ - E$_{\rm{pk}}$ correlation for the sample of 81 long GRBs (\textit{black diamonds} for the old sample, \textit{blue dots} for the new sample, \textit{green triangles} for low-energy GRBs, and \textit{cyan triangles} for the ultra-long GRBs), 11  short GRBs (\textit{red pentagons} for the old sample and \textit{red stars} for the new sample), and two GRBs with uncertain classification (\textit{yellow squares}). The \textit{black solid line} is the best-fitting function Log[E$_{\rm{x,iso}}$] = 0.96 Log[E$_{\rm{\gamma,iso}}$] - 0.57 Log[E$_{\rm{pk}}$] - 0.62 and the \textit{yellow area} marks the 2$\sigma$ confidence region. The \textit{gray area} indicates the position of the intermediate group.}
\label{fig:3par}
\end{figure*}


\section{The E$_{\rm{X,iso}}$-E$_{\rm{\gamma,iso}}$-E$_{\rm{pk}}$ correlation}
\label{sec:three}

The E$_{\rm{X,iso}}$-E$_{\rm{\gamma,iso}}$-E$_{\rm{pk}}$ correlation involves E$_{\rm{pk}}$, E$_{\rm{\gamma, iso}}$, and E$_{\rm{X,iso}}$. E$_{\rm{pk}}$ and E$_{\rm{\gamma, iso}}$ are calculated as described in \citet{2002A&A...390...81A}, where the gamma-ray spectrum is fitted using a Band function \citep{1993ApJ...413..281B}. For the short GRBs 080123, 090423, 100625A, 111117A, and 130603B, we consider the values of E$_{\rm{pk}}$ and E$_{\rm{\gamma, iso}}$ reported in \citet{2014MNRAS.442.2342D}, who fitted the spectrum considering also a cut-off power-law function or a Band function with fixed high-energy index.

For GRBs in the original sample, we extracted the XRT light-curves using the method presented in \citet{Margutti09}, while for GRBs observed after December 2010, we used the count-rate light-curves from the official repository site  \citep{2007A&A...469..379E,2009MNRAS.397.1177E} and employed the time-resolved spectral analysis to perform the spectral calibration in the common rest-frame energy band 0.3-30 keV. In both cases we use a time-variable flux-to-count conversion factor and we propagate the uncertainty from the spectral fits to the final flux light-curves. This method allows us to correctly detect the presence of statistically significant positive temporal fluctuations superimposed on a smoothly decaying light-curve. Finally, E$_{\rm{X,iso}}$ was calculated as in M13, fitting the continuum part of  0.3 - 30 keV X-ray light-curves in luminosity (\citealt{2011MNRAS.417.2144M}; M13; \citealt{zaninoni13}) and integrating the fitted light-curve between the start and the end time of the observations. We compute the E$_{\rm{X,iso}}$ in the 0.3 - 30 keV band for considering the bulk of the X-ray emission. On the other hand, as we demonstrate in B12 and M13, the choise of the 0.3- 30 rest frame band or of the 0.3 -10 keV observer band does not influence the reliability of the correlation.

In Table \ref{tab:sampletot} in Appendix we listed the values of E$_{\rm{X,iso}}$, E$_{\rm{\gamma,iso}}$ and E$_{\rm{pk}}$ for the new GRBs of our sample.

The correlation is derived using the method of \citet{2005physics..11182D} (for details see Appendix \ref{agostini}), which considers an intrinsic scatter $\sigma_{\rm{ext}}$ that accounts for the possible contribution of hidden variables\footnote{For this analysis we use a procedure that uses \textit{R} (http://www.r-project.org/) program language.} (see also B12, M13).
In this way:

\begin{eqnarray}
\rm{Log}\left[\frac{E_{\rm{X,iso}}}{\rm{erg}}\right]& =& (0.97 \pm 0.06)\ \rm{ Log}\left[\frac{E_{\rm{\gamma,iso}}}{\rm{erg}}\right]\\
&&-(0.57\pm0.13)\ \rm{Log}\left[\frac{E_{\rm{pk}}}{\rm{keV}}\right] - (0.62\pm0.08), \nonumber
\end{eqnarray}

with an extra-scatter $\sigma_{\rm{ext}}$ = 0.32 $\pm$ 0.04. Figure \ref{fig:3par} shows a two-dimensional representation of this relation. The extra-scatter is similar to the value obtained with the old sample (B12, M13). As explained in B12 and M13, this correlation is robust, spanning four orders of magnitude in E$_{\rm{X,iso}}$ and E$_{\rm{pk}}$, and six orders of magnitude in E$_{\rm{\gamma,iso}}$, and combines both short and long GRBs in a common scaling. Indeed, the newly added short GRBs (080123, 100117A, 100625A, 111117A, 130603B; Table \ref{tab:sampleshort}) confirm the validity of the E$_{\rm{X,iso}}$-E$_{\rm{\gamma,iso}}$-E$_{\rm{pk}}$ correlation for all families of GRBs and populate the same region of the plane as short GRBs of the original sample.

If we consider only long GRBs the relation becomes:

\begin{eqnarray}
\rm{Log}\left[\frac{E_{\rm{X,iso}}^{\rm{L}}}{\rm{erg}}\right]& =& (0.77 \pm 0.14)\ \rm{ Log}\left[\frac{E_{\rm{\gamma,iso}}^{\rm{L}}}{\rm{erg}}\right]\\
&&-(0.21\pm0.24)\ \rm{Log}\left[\frac{E_{\rm{pk}}^{\rm{L}}}{\rm{keV}}\right] - (0.63\pm0.08), \nonumber
\end{eqnarray}
with an extra-scatter $\sigma_{\rm{ext}}$ = 0.31$\pm$0.04. The extra-scatter is similar to that of the entire sample, while the slope is slightly different.

We test the  E$_{\rm{X,iso}}$-E$_{\rm{\gamma,iso}}$-E$_{\rm{pk}}$ correlation computing the correlation coefficient ($\rho_{xy}$) and the null hypothesis probability (NHP) for the new and the old sample. In the case of the complete samples we obtain $\rho_{xy}^{\rm{new}}$ = 0.92 and NHP$_{\rm{new}} \sim$ 1 for the updated sample (94 GRBs) and $\rho_{xy}^{\rm{old}}$ = 0.90 and NHP$^{\rm{old}} \sim$ 1 for the old sample (61 GRBs); if we consider only long GRBs, we obtain $\rho_{xy}^{\rm{new,long}}$ = 0.86 and NHP$_{\rm{new,long}} \sim$ 1 for the updated sample (81 GRBs) and $\rho_{xy}^{\rm{old}}$ = 0.84 and NHP$_{\rm{old,long}} \sim$ 1 for the old sample (54 GRBs). Therefore the existence of this correlation is confirmed and, since $\rho_{xy}$ and NHP are larger for the updated sample than the old one, we can conclude that the correlation is stronger.

\begin{table*}
\caption{Short GRBs. \textit{GRB}: GRB name. $z$: redshift. $E_{\rm{pk}}$: peak energy in keV units. \textit{Notes}: it indicates if the GRB belongs to the original sample ($OS$) or to the new sample ($NS$) and if it was used for the fit of the E$_{\rm{X,iso}}$-E$_{\rm{\gamma,iso}}$-E$_{\rm{pk}}$ correlation ($Y$) or not ($N$). \textit{z Ref.} and \textit{E$_{\rm{pk}}$ Ref.}: reference for the redshift and the E$_{\rm{pk}}$, respectively. (1) \citet{2013MNRAS.428..729M}; (2) \citet{2014MNRAS.442.2342D}; (3) \citet{2013ApJ...765..121B}; (4)  \citet{2012ApJ...756...63M}; (5) \citet{2005GCN..4403....1S}; (6). \citet{2008MNRAS.391..577A}.}
\label{tab:sampleshort}
\begin{tabular}{lllcccl}
\hline
GRB & $z$ & E$_{\rm{pk}}$ (keV) & Notes & & z Ref. & E$_{\rm{pk}}$ Ref. \\
\hline
050724  & 0.258  & 100$\pm$16        & OS & Y & (1) & (1) \\
051221A & 0.5465 & 622$\pm$35        & OS & Y & (1) & (1) \\
051227  & 0.714  & 100$^{+219}_{-41.3}$ & NS &  N  & (1) & (5) \\
061006  & 0.438  & 955$\pm$259       & OS & Y & (1) & (1) \\
061201  & 0.111  & 969$\pm$412       & NS & N & (6) & (6)\\
070714B & 0.92   & 215$\pm$750       & OS & Y & (1) & (1) \\
071227  & 0.3830 & 1384$\pm$277      & OS & Y & (1) & (1) \\
080123  & 0.495  & 149.50            & NS & Y & (2) & (2) \\
090510  & 0.903  & 8370$\pm$760      & OS & Y & (1) & (1) \\
100117A & 0.92   & 551$\pm$135       & NS & Y & (1) & (1) \\
100625A & 0.452  & 701.32$\pm$114.71 & NS & Y & (2) & (2) \\
101219A & 0.718  & 842$\pm$170       & NS & N & (2) & (2) \\
111117A & 1.2    & 966$\pm$322       & NS & Y & (4) & (2) \\
120804A & 1.3    & 310               & NS & N & (3) & (3) \\
130603B & 0.356  & 900$\pm$140       & NS & Y & (2) & (2) \\
\hline
\end{tabular}
\end{table*}


\subsection{Ultra-long GRBs}
Recently, there has been extensive discussion about the existence of a new class of GRBs, named ultra-long GRBs (e.g. \citealt{2011Natur.480...72T, 2012ApJ...748...59G, 2013ApJ...766...30G, 2013ApJ...778...54V, 2014ApJ...787...66Z, 2014ApJ...781...13L}). From the point of view of the duration of the prompt emission of GRBs, some authors classify as ultra-long GRBs those bursts with durations of several thousand seconds (e.g. \citealt{2014ApJ...781...13L}), while other authors give a more precise definition: long GRBs have T$_{90} >$ 2 s, very-long GRBs have T$_{90} >$ 10$^3$ s, while ultra-long GRBs have T$_{90} >$ 10$^4$ s (e.g. \citealt{2013ApJ...766...30G})\footnote{T$_{90}$ is the time in which from 5\% to 95\% cumulative counts are recorded.}. Only a few very- and ultra- long GRBs were observed so far. Among these \textit{Swift} detected low-energy GRBs\footnote{We consider as low-energy GRBs long GRBs with E$_{\rm{\gamma,iso}}$ below 10$^{52}$ erg.} 060218 (T$_{90}\ =$ 2100 s; \citealt{2006GCN..4775....1C}) and 100316D (T$_{90}>$1300 s; \citealt{2010GCN..10496...1S}), and GRBs 101225A (T$_{90}\geq$1650 s; \citealt{2010GCN..11493...1R}), 111209A (T$_{90}\sim$15000 s; \citealt{2011GCN..12632...1H}), and 130925A (T$_{90}\sim$ 20000 s, \citealt{2013GCN..15246...1L}). 

In our sample there are three ultra long lasting bursts: low-energy GRB 060218 and GRBs 111209A and 130925A. In what follows we consider also GRB 101225A, which was not considered in our previous sample because of the uncertainty of its redshift \citep{2011Natur.480...69C, 2011Natur.480...72T}, now settled to be 0.847 \citep{2014ApJ...781...13L}. Ultra-long GRBs do not occupy a particular area in the E$_{\rm{\gamma, iso}}$ - E$_{\rm{pk}}$ - E$_{\rm{X,iso}}$ plane (Figure \ref{fig:3par} \textit{cyan triangles}). In particular, GRB 060218 is a low-energy GRB and lies consistently in the bottom-left area of this plane, while GRB 111209A and GRB 130925A lie in the region occupied by other long GRBs. GRB 101225A follows the E$_{\rm{X,iso}}$-E$_{\rm{\gamma,iso}}$-E$_{\rm{pk}}$ correlation and behaves as a long GRB. When we consider other 2-parameter correlations E$_{\rm{\gamma,iso}}$ - E$_{\rm{pk}}$ (Figure \ref{fig:amati}), E$_{\rm{pk}}$ - E$_{\rm{X, iso}}$ (Figure \ref{fig:exepk}), E$_{\rm{\gamma,iso}}$ - E$_{\rm{X, iso}}$ (Figure \ref{fig:exegamma}), and E$_{\rm{pk}}$ - $\epsilon$ ($\epsilon=1/\eta$ with $\eta$ the efficiency of the process, see Sec. \ref{sec:physic} for more details; Figure \ref{fig:efficiency}), we notice that GRBs 101225A, 111209A, and 130925A behave like ordinary long GRBs, while GRB 060218 lies in a peculiar area because of its low energies both in the X-rays and $\gamma$-rays\footnote{GRB 101225A does not lie in the 2-sigma region for E$_{\rm{pk}}$ - $\epsilon$ relation. The uncertainties about the analysis of the data of this GRB prevent us from deriving firm conclusions about its behavior.}. 

Ultra-long GRBs are consistent with the E$_{\rm{X,iso}}$-E$_{\rm{\gamma,iso}}$-E$_{\rm{pk}}$ correlation as all other GRBs. Even if these GRBs seem to show uncommon local properties (e.g. \citealt{2014ApJ...781...13L}), their similar behavior in the E$_{\rm{X,iso}}$-E$_{\rm{\gamma,iso}}$-E$_{\rm{pk}}$ correlation could be related to a general and not local feature, as the dynamics of the jet.



\subsection{GRBs with uncertain classification}
\label{sec:threethree}
GRBs 090426 and 100816A have an uncertain classification, because they show properties that are intermediate between long and short GRBs. From Figure \ref{fig:3par}, we note that short GRBs occupy the left bottom part of the correlation plane, instead GRB 090426 lies between the groups of long and short GRBs. Indeed the classification of this GRB is very debated (e.g., \citealt{2009A&A...507L..45A, 2010MNRAS.401..963L, 2011MNRAS.414..479T, 2011A&A...531L...6N, 2013ApJS..209...20G, 2014MNRAS.442.2342D}). Even if GRB 090624 has a very soft spectral index and it lies in the 2$\sigma$ confidence level region of the Amati relation, we consider this burst as a short GRB, as discussed also by other authors (e.g., \citealt{2014MNRAS.442.2342D}). Its gamma-ray emission duration is less than 2 s (T$_{90}^{\rm{obs}} <$0.2 s and T$_{90}^{\rm{RF}}<0.5$ s). In addition, the value of the intrinsic absorption (N$_{\rm{H}} = 2.3^{+5.6}_{-19}\times10^{21}$ cm$^{-2}$), the presence of the highly ionized absorption lines, the position of the afterglow in the host galaxy, and the duration of the gamma-ray emission, support the idea that GRB 090426 was formed by the merger of two compact objects. In some scenarios (e.g. \citealt{2006ApJ...648.1110B,2002ApJ...570..252P}), the duration of the merger is very short and the binary system remains inside the star forming region, as observed in this case, so the afterglow luminosity could be comparable to that of long GRBs. This scenario supports also the high redshift of GRB 090426.

GRB 100816A lies between long and short GRB groups in the E$_{\rm{X,iso}}$-E$_{\rm{\gamma,iso}}$-E$_{\rm{pk}}$ correlation and its classification is still uncertain \citep{2012IAUS..279..415T,2011ApJ...735...23N,2013MNRAS.428..729M,2014MNRAS.442.2342D}. It can be classified as short GRB because it has an hard spectrum, it lies offset from its host galaxy, and there is no association with a supernova (SN). On the other hand, in strict analogy to long GRBs, it follows the Amati relation, its gamma-ray duration is longer than 2 s (T$_{90}\sim$ 2.9 s), and has a positive spectral lag \citep{2015MNRAS.446.1129B}.


\section{Discussion}
\label{sec:discussion}
The new sample of GRBs confirms the existence of the E$_{\rm{X,iso}}$-E$_{\rm{\gamma,iso}}$-E$_{\rm{pk}}$ correlation, with similar best-fitting parameters. In particular, the larger number of short GRBs better constrains this correlation, since there are more bursts with low energies. In addition, this correlation is robust, spanning four orders of magnitude in E$_{\rm{X,iso}}$ and E$_{\rm{pk}}$, and six orders of magnitude in E$_{\rm{\gamma,iso}}$, and combining both short and long GRBs in a common scaling. In the following sections we discuss the presence of an \textit{intermediate} group of GRBs that lies between long and short GRBs in the E$_{\rm{X,iso}}$-E$_{\rm{\gamma,iso}}$-E$_{\rm{pk}}$ correlation plane (Section \ref{sec:inter}), the possible presence of biases that shape the correlation (Section \ref{sec:bias}), and the possible physical processes and mechanisms that lead to the E$_{\rm{X,iso}}$-E$_{\rm{\gamma,iso}}$-E$_{\rm{pk}}$ correlation (Section \ref{sec:physic}).


\subsection{The intermediate group}
\label{sec:inter}
In the previous compilation of the E$_{\rm{X,iso}}$-E$_{\rm{\gamma,iso}}$-E$_{\rm{pk}}$ correlation, GRBs were divided in two groups along the correlation with a lack of objects between them. Thanks to the updated sample, this area is now occupied by three new long GRBs (110106B, 120724A, 130831A) and a GRB with uncertain classification (090426). This group of objects, together with short GRBs 070714B and 090510, low-energy GRBs 050416A, 060614, and 081007, long GRB 080916, and GRB 100816A with uncertain classification, suggest the presence of an intermediate group between long and short GRBs (Figure \ref{fig:3par}, \textit{gray area}). 

In this group there are not only the two GRBs with uncertain classification, of which we discussed in Sec. \ref{sec:threethree}, but also another intriguing object, GRB 060614. This GRB has been classified as long GRB because of the duration of its gamma-ray emission (T$_{90}$ = 102 s), but it has some characteristics that are typical of short GRBs (e.g. \citealt{2006Natur.444.1044G,2006Natur.444.1047F,2006Natur.444.1050D,2006Natur.444.1053G}). In spite of its low redshift ($z$ = 0.125), no SN has been detected and its environment is typical of short GRBs, since it exploded in a zone with a small specific star formation rate and off-set from the GRB host nucleus. On the other hand, its X-ray, UV and optical light-curves and spectral energy distributions (SEDs) are well explained within the standard afterglow model \citep{2007A&A...470..105M, 2009ApJ...703...60X} and it follows the Amati relation as a normal long burst \citep{2007A&A...463..913A}. Because of the peculiarity of this object alternatives progenitors has been proposed, as a compact binary merger or a massive collapsar that powers a GRB with no association with a SN (e.g. \citealt{2006Natur.444.1044G, 2006Natur.444.1047F,   2006Natur.444.1050D, 2006Natur.444.1053G, 2007ApJ...655L..25Z, 2009A&A...498..501C}).

The \textit{Swift}-BAT light-curve of the short GRB 070714B shows a short-duration peak followed by a softer, long-lasting tail, called extended emission (EE). One quarter of the detected short GRBs have an EE \citep{2010ApJ...717..411N}. Their gamma-ray light-curves are similar and uniform each other and their X-ray light-curves have similar plateau luminosities and time-scales \citep{2013MNRAS.431.1745G}. This suggests that they have a common progenitor, which is different from the standard merger scenario for short GRBs, since it must require an injection of energy after the first spike, which switches off around 100 s after the trigger in the rest frame. Several models have been proposed to explain the origin of these objects; for example, the central engine for these bursts could be a magnetar (e.g. \citealt{2008MNRAS.385.1455M, 2012MNRAS.419.1537B, 2013MNRAS.431.1745G}).


\begin{figure}
\centering
\includegraphics[width=\columnwidth]{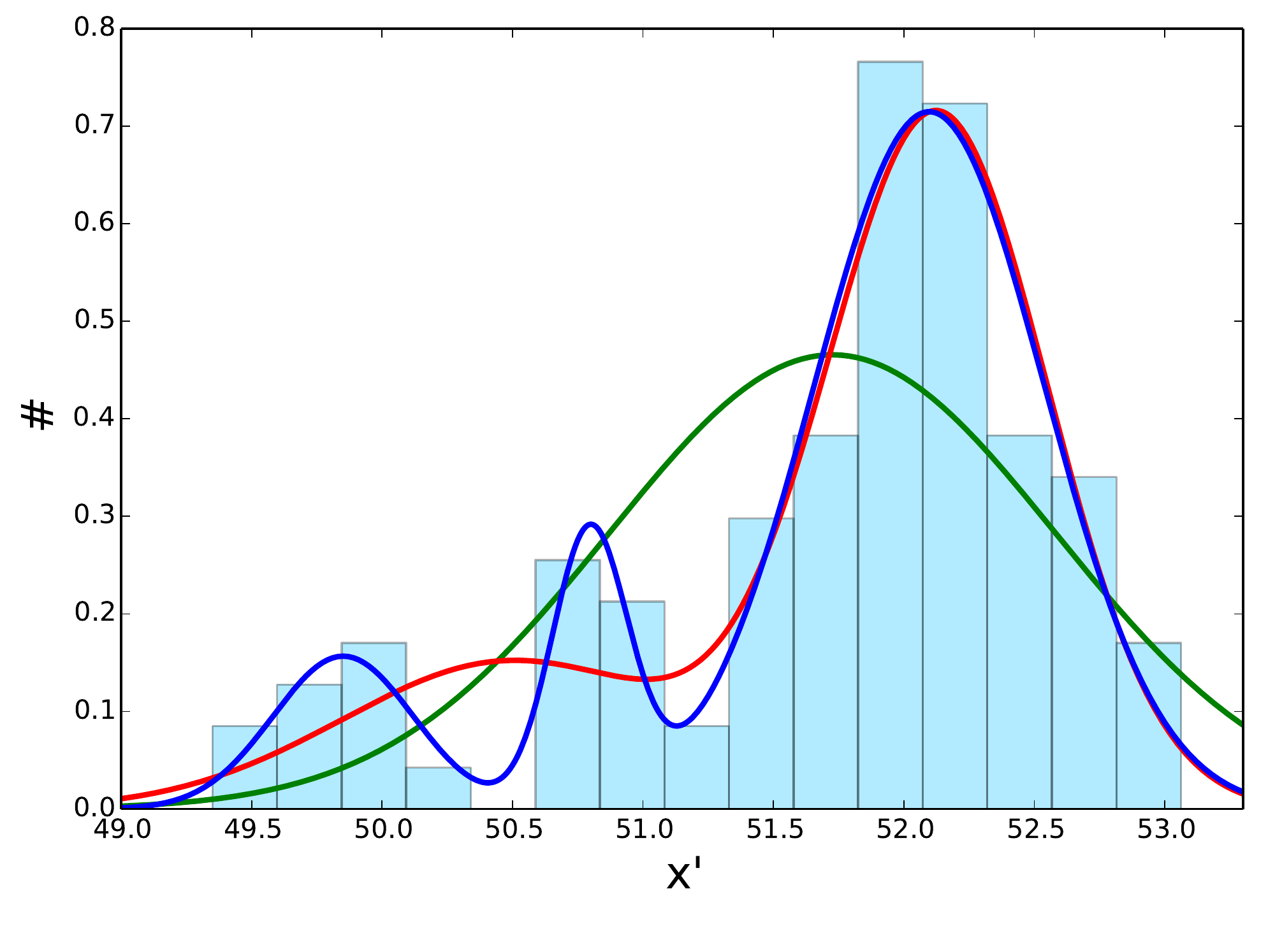}
\caption{Distribution of GRBs in our sample as projected over the best fit function of the E$_{\rm{X,iso}}$-E$_{\rm{\gamma,iso}}$-E$_{\rm{pk}}$ correlation, with $x'$ the coordinate representing the projection of the data over this function. We fit the distribution with a Gaussian function (\textit{green solid line}), the sum of two Gaussian functions (\textit{red solid line}), and the sum of three Gaussian functions (\textit{blue solid line}).}
\label{fig:histo}
\end{figure}

\begin{figure*}
\centering
\subfloat[][]{\includegraphics[width=0.45\textwidth]{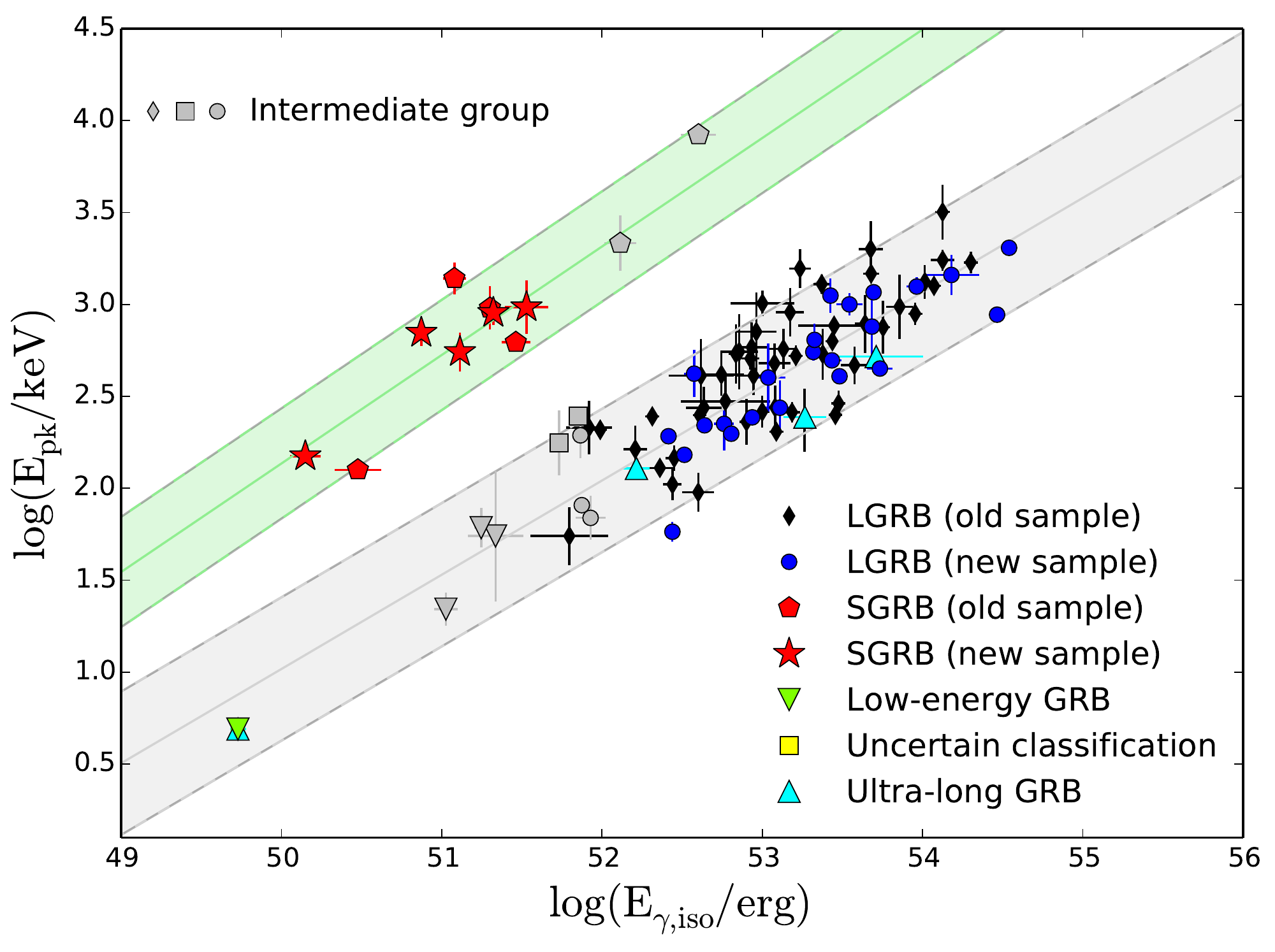}\label{fig:amati}}
\qquad
\subfloat[][]{\includegraphics[width=0.45\textwidth]{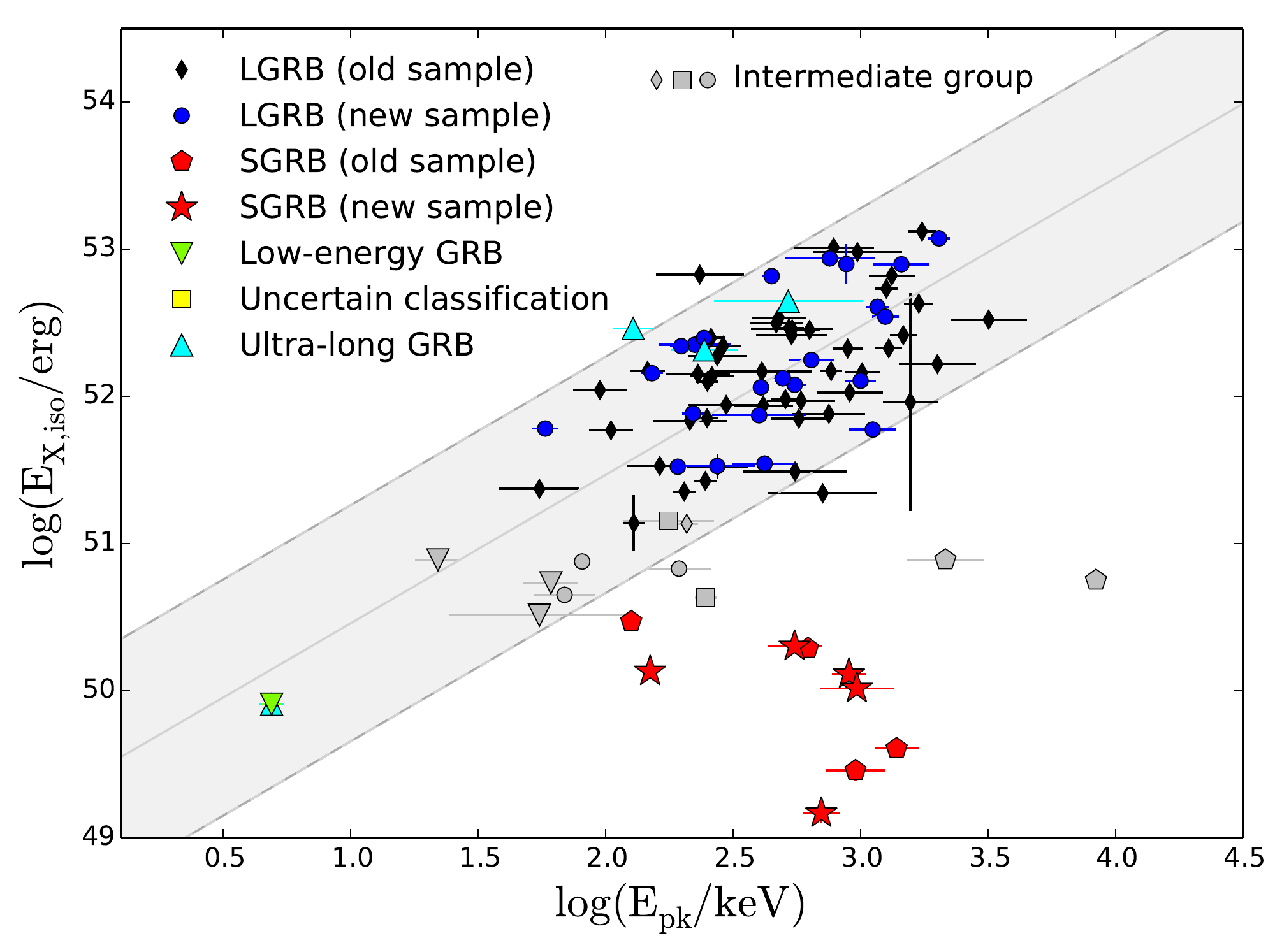}\label{fig:exepk}}
\qquad
\subfloat[][]{\includegraphics[width=0.45\textwidth]{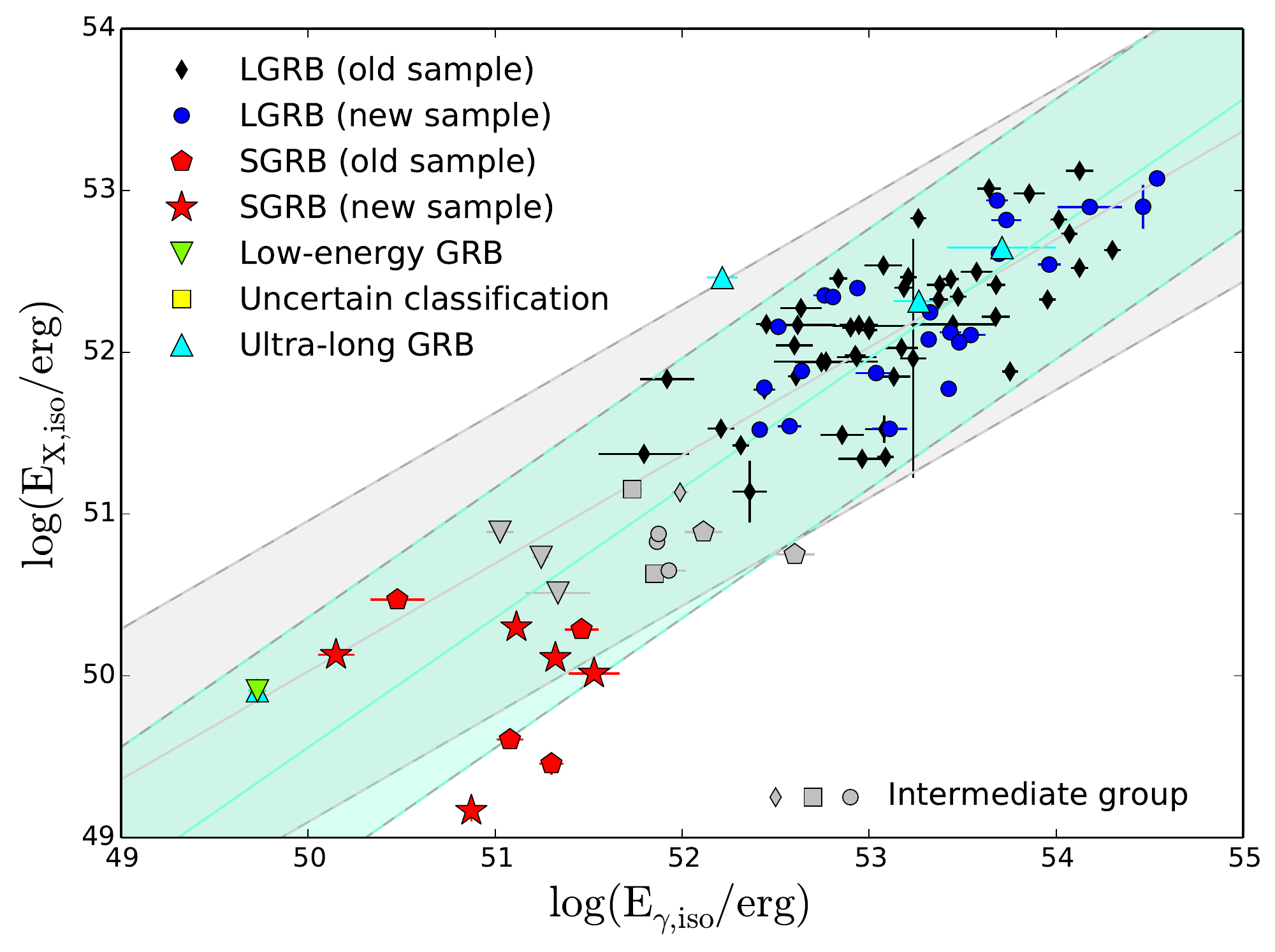}\label{fig:exegamma}}
\qquad
\subfloat[][]{\includegraphics[width=0.45\textwidth]{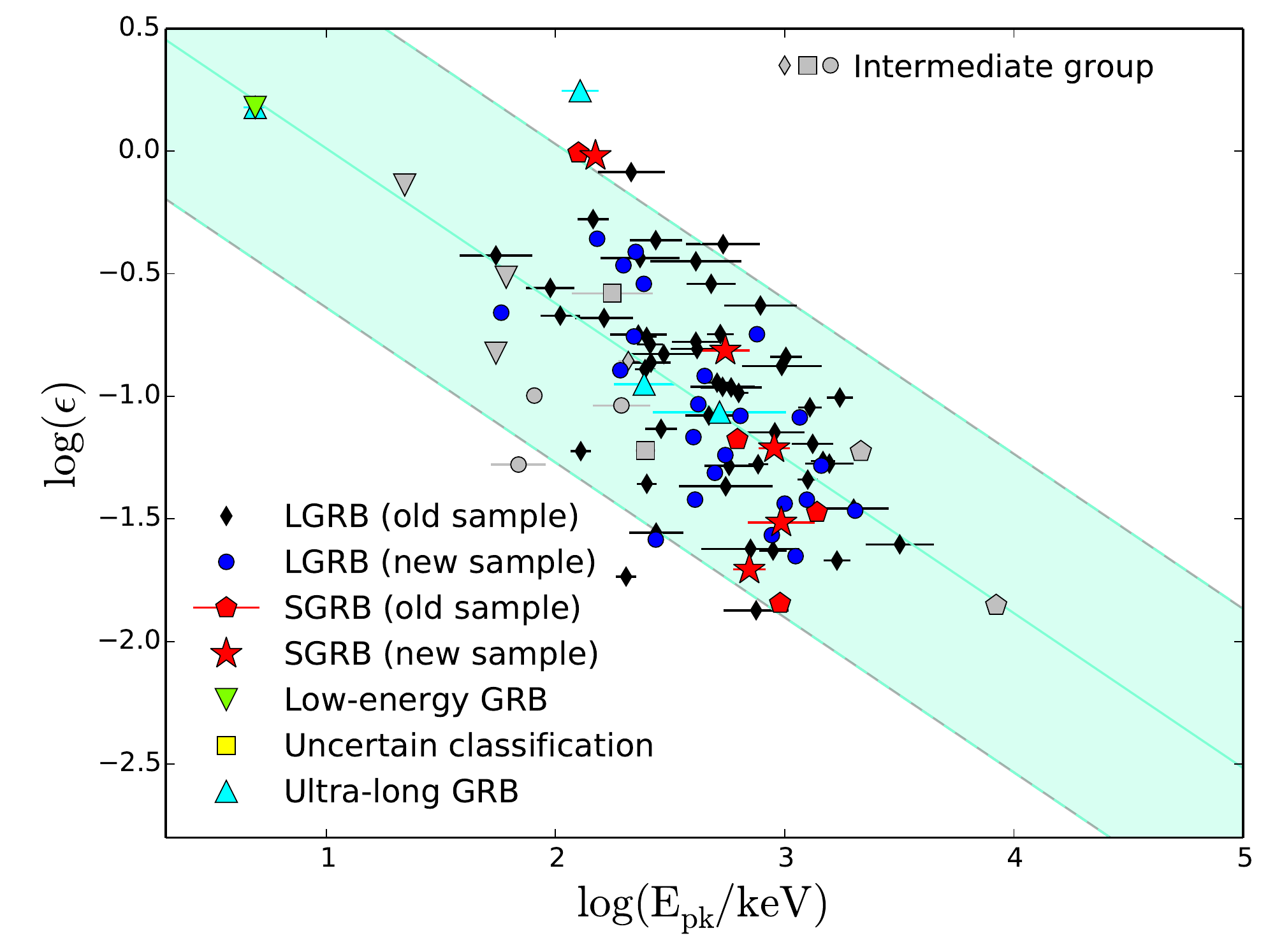}\label{fig:efficiency}}
\caption{Two-parameter relations.  Color code as Figure \ref{fig:3par}. \textit{Gray} symbols indicate the intermediate group. The \textit{gray} area indicates that the best fit is computed using only long GRBs, while the \textit{cyan} area indicates that the best fit is computed using all GRBs of the sample. (a) The Amati relation (E$_{\rm{\gamma, iso}}$ - E$_{\rm{pk}}$ relation): the \textit{green} solid line is the best fit function for short GRBs as calculated by \citet{2015MNRAS.448..403C} and the \textit{green} area marks the 2$\sigma$ region. (b) E$_{\rm{x,iso}}$ - E$_{\rm{pk}}$ relation. (c) E$_{\rm{\gamma, iso}}$ - E$_{\rm{x,iso}}$ relation. (d) E$_{\rm{pk}}$ vs. $\epsilon$.}

\end{figure*}
Figure \ref{fig:histo} shows the distribution of GRBs in our sample as projected over the best fitting function of the E$_{\rm{X,iso}}$-E$_{\rm{\gamma,iso}}$-E$_{\rm{pk}}$ correlation, with $x'$ the coordinate representing the position of the data points over this new reference axis (e.g. the best fit function)\footnote{If, e.g., P($x_{\rm P}$; $y_{\rm P}$) is the data point and $y = m x + q$ is our best fit function, the coordinate $x'$ is the $x$-coordinate of the intersection point between the line that passes for the point P and perpendicular to the best fit function, and the best fit function itself. In this way, $x' = (m / (m^2 + 1)) (y_{\rm P} - q + (x_{\rm P} / m))$.}. Following the procedure used by \citet{2008A&A...489L...1H}, we fit the distribution with a single Gaussian function ($G1$), with the sum of two Gaussian functions ($G2$), and then with the sum of three Gaussian functions ($G3$) using the Maximum Likelihood method\footnote{For this analysis we minimize the negative logarithm of the Likelihood function using the procedure \texttt{scipy.optimize.minimize} of Python program language (https://www.python.org/). For the three fit functions considered we obtain: $a)$ single Gaussian function: $\mu$ = 51.73 and $\sigma$ = 0.86; $b)$ sum of two Gaussian functions: $\mu_1$ =  50.50, $\sigma_1$ = 0.65, $\mu_2$ = 52.13 and $\sigma_2$ = 0.42; $c)$ sum of three Gaussian functions: $\mu_1$ = 49.85, $\sigma_1$ = 0.27, $\mu_2$ = 50.80, $\sigma_2$ = 0.14, $\mu_3$ = 52.10, and $\sigma_3$ = 0.44.}. The maximum value of the log-likelihood functions are respectively, -120.12, -104.58, -100.41. We perform a likelihood ratio (LR) test\footnote{We define LR = 2 [ln(L1) - ln(L2)], with L1 and L2 the maximum log-likelihood for the simpler model and the more complicated model, respectively. Since LR $\sim\ \chi^2$, with the degrees of freedom (DOF) equal to the number of additional parameters in the more complex model, we can calculate the p-value. If p-value $<$ 0.05 the more complex model is favorable.} to verify which model fits better this distribution and we obtain: LR$_{G1,G2}$ = 31.08 with p-value = 0.01 (DOF = 3),  LR$_{G2,G3}$ = 8.34  with p-value = 0.04 (DOF = 3) and  LR$_{G1,G3}$ = 40 with p-value = 0 (DOF = 6). From this statistical analysis, we can affirm that the best fit is done with the model $G3$.

In the Amati relation (Figure \ref{fig:amati}, Table \ref{tab:relations}) the intermediate group lies in the low peak energy part of the plane, with the exception of two short GRBs, since the Amati relation is followed only by long GRBs. GRB 090426 and GRB 100816A are within in the limit of 2$\sigma$ of the Amati relation, so they behave as long GRBs \citep{2014MNRAS.442.2342D}. 

In the E$_{\rm{pk}}$ - E$_{\rm{X,iso}}$ relation for long GRBs (Figure \ref{fig:exepk}, Table \ref{tab:relations}), GRBs of the intermediate group lie within 2$\sigma$ from the best fitting function, with exception of GRB 100816A wich falls into 3$\sigma$ and short GRBs do not follow this relation.

The E$_{\rm{\gamma,iso}}$ - E$_{\rm{X,iso}}$ relation is followed by all GRBs of the intermediate group (Figure \ref{fig:exegamma}, \textit{cyan} and \textit{gray areas}, respectively; Table \ref{tab:relations}).

The region of the E$_{\rm{X,iso}}$-E$_{\rm{\gamma,iso}}$-E$_{\rm{pk}}$ correlation plane occupied by this group could be considered as an intermediate zone. In this area we find both low- and high-energy GRBs. These objects have similar X-ray energies around 10$^{51}$ erg, but very different peak energies. The isotropic X-ray energy of both short and long GRBs linearly increases with the isotropic gamma-ray energy with the same scaling law (Figure \ref{fig:exegamma}), while, for a given isotropic gamma-ray energy or a fixed X-ray energy, short GRBs have higher peak energies than long GRBs (Figures \ref{fig:amati} \ref{fig:exepk}).

In addition, the majority of GRBs in the intermediate group has a redshift $<$ 1, except for short GRB 090426 ($z$ =  2.609) and GRB 120724A ($z$ = 1.48), and long GRBs belonging to this group have typical durations (T$_{90} \leq $ 80 s, GRB 070714B). Their X-ray light-curves have a canonical shape, with an initial steep decay, followed by a plateau of small duration and/or steeper than usual and a normal decay phase\footnote{The X-ray light-curves of GRBs 050416A and 100816A show a small steep decay, difficult to fit with a canonical shape, while the observations of GRB 120724A stop before 24000 s after the trigger, suggesting that the normal decay phase might have been missing.}.  

Indeed these GRBs have small redshifts, like short GRBs, and are less energetic than long one, even if they have typical long GRB durations. They have canonical X-ray light-curves and they show limited flaring activity. Therefore they seem to represent a transitional group between GRBs with low energies and simple X-ray light-curves (e.g. single or double power laws or canonical shapes) and more energetic long GRBs, with also complex and unusual X-ray light-curves (e.g. with a shallow phase before the steep decay or with big flares). A larger sample is needed to confirm or rule out the presence of an intermediate group.


\subsection{Possible biases}
\label{sec:bias}
In B12 and M13 we discussed the possible caveat on the definition of E$_{\rm{X,iso}}$. In particular, we analysed the differences between the values computed in the observer frame 0.3-10 keV and in the rest frame 0.3-30 keV, and on the arbitrariness of the choice of the interval time for the integration. We concluded that these factors do not influence the E$_{\rm{X,iso}}$-E$_{\rm{\gamma,iso}}$-E$_{\rm{pk}}$ correlation. The E$_{\rm{X,iso}}$ does not include the contribution of flares, which are present in $\sim$ 40\% of light-curves of our sample. As we discussed in B12 and M13, the inclusion of flares in the computation of the E$_{\rm{X,iso}}$ does not influence the  E$_{\rm{X,iso}}$-E$_{\rm{\gamma, iso}}$-E$_{\rm{pk}}$ correlation because the energy content of flares is usually $\sim$ 25\% of  the underlying continuum E$_{\rm{X, iso}}$ and the correlation scatter does not better, since the most scattered population in this correlation are short GRBs that have no bright flares \citep{2011MNRAS.417.2144M}.

In this Section, we examine if the E$_{\rm{X,iso}}$-E$_{\rm{\gamma,iso}}$-E$_{\rm{pk}}$ correlation could be affected by the redshift and by the properties of the X-ray light-curve.

Regarding the relation between E$_{\rm{X,iso}}$ and redshift, in Figure 4 in M13 we showed that we are not sensitive to the population of bursts with E$_{\rm{X, iso}}$ $<$ 10$^{51}$ erg for $z$ $>$ 2, in this way the low-energy tail of the E$_{\rm{X, iso}}$ distribution is currently undersampled. This is likely a non-detectability zone. For $z$ $>$ 1 there is no evidence for an evolution of the upper bound of E$_{\rm{X,iso}}$ with redshift, which may suggest that $\sim$10$^{53}$ erg is a physical boundary to the E$_{\rm{X, iso}}$ distribution. M13 underlined that maximum budget E$_{\rm{max}}\ \sim\ 10^{52}$ erg is predicted by magnetar models \citep{1992Natur.357..472U}.

\begin{figure}
\centering
\includegraphics[width=\columnwidth]{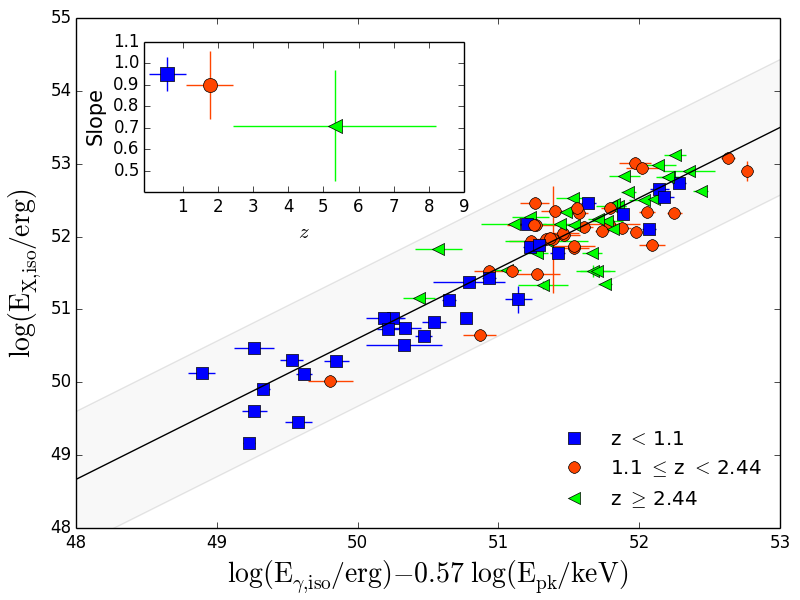}
\caption{E$_{\rm{X,iso}}$-E$_{\rm{\gamma,iso}}$-E$_{\rm{pk}}$ correlation, considering the differences in redshift for GRBs in our sample. \textit{Blue squares}: $z$ $<$ 1.1; \textit{orange dots}: 1.1 $\leq$ $z$ $<$ 2.44; \textit{green triangles}: $z$ $\geq$ 2.44.  The \textit{black solid line} is the best fit function of the complete sample and the \textit{gray area} marks the 2$\sigma$ regions. \textit{Inset}: slopes of the best fit function for the three groups.}
\label{fig:redshift}
\end{figure}

In Figure \ref{fig:redshift}, we show the distribution of GRBs in the E$_{\rm{\gamma,iso}}$ - E$_{\rm{pk}}$ - E$_{\rm{X,iso}}$ plane depending on their redshift $z$. Low X-ray energy GRBs (short GRBs and low-energy long GRBs) are observed only at low redshift, while long GRBs are observed at every redshift. We divided our sample into three groups with the same number of objects depending on their redshift ($z$ $<$ 1.1, 31 GRBs; 1.1 $\leq$ $z$ $<$ 2.44, 32 GRBs; $z$ $\geq$ 2.44, 32 GRBs) and we calculated the best fit function for each group\footnote{Best fit function for the three groups of GRBs: $a)$ $z$ $<$ 1.1, $\rm{Log}\left[E_{\rm{X,iso}}\right] = (0.95 \pm 0.08)\ \rm{Log}\left[E_{\rm{\gamma,iso}}\right]-(0.61\pm0.14)\ \rm{Log}\left[E_{\rm{pk}}\right] - (0.66\pm0.10)$, with $\sigma_{\rm{ext}}= 0.29\pm0.06$; $b)$ 1.1 $\leq$ $z$ $<$ 2.44, $\rm{Log}\left[E_{\rm{X,iso}}\right] = (0.90 \pm 0.16)\ \rm{ Log}\left[E_{\rm{\gamma,iso}}\right]-(0.47\pm0.31)\ \rm{Log}\left[E_{\rm{pk}}\right] - (0.61\pm0.19)$, with $\sigma_{\rm{ext}}= 0.32\pm0.06$; $c)$ $z$ $\geq$ 2.44, $\rm{Log}\left[E_{\rm{X,iso}}\right] = (0.71 \pm 0.26)\ \rm{ Log}\left[E_{\rm{\gamma,iso}}\right]-(0.24\pm0.46)\ \rm{Log}\left[E_{\rm{pk}}\right] - (0.49\pm0.21)$, with $\sigma_{\rm{ext}}= 0.35\pm0.07$.}; as shown in Figure \ref{fig:redshift} (\textit{Inset}), the slope of the correlation does not evolve with $z$. The error of the slope parameter increases with the $z$, maybe because of the lack of high-$z$ objects in the bottom left part of the plane (i.e. low-energy GRBs)\footnote{To prove that the E$_{\rm{\gamma,iso}}$ - E$_{\rm{pk}}$ - E$_{\rm{X,iso}}$ relation does not depend on $z$ a further analysis is needed, for example as in \citet{2013ApJ...774..157D} for the $L_{X}$ - $T_{a}^{\star}$ relation, but it is beyond the aim of this paper.}.

Moreover, the estimation of E$_{\rm{X,iso}}$, E$_{\rm{\gamma,iso}}$, and E$_{\rm{pk}}$ can be influenced by different factors, for example the systematics introduced by the limited energy band of the detector (e.g. \citealt{2000ApJ...534..227L, 2002ApJ...565..182L}), the extrapolation made for computing the energies in an rest frame energy band (e.g. \citealt{2013ApJ...765..116K}), and the choice of the cosmological parameters to calculate the luminosity distance (\citealt{2006MNRAS.372..233A}). On the other hand, since the slope of this correlation is $\sim$ 1, the dependence from the luminosity distance is small.

In particular, analysing E$_{\rm{\gamma,iso}}$ vs. $z$, we note that there are few very bright GRBs at low redshift: this could be caused by evolutionary effects (i.e. older GRBs are brighter), by a correlation between GRB brightness and jet opening angle (e.g. \citealt{2013MNRAS.428.1410G}), or by a combination of jet structure and viewing angle (e.g. \citealt{2013ApJ...765..103L}). For E$_{\rm{pk}}$, it is necessary to underline that the maximum detectable E$_{\rm{pk}}$ depends on the combination between the brightness of the GRBs and the effective collecting area of the instrument respect to the energy band.  Because of these effects, the correlation could be slightly influenced by $z$, but current data do not allow to confirm this issue (see for example Figure \ref{fig:redshift}, \textit{Inset}).

\begin{figure}
\centering
\includegraphics[width=\columnwidth]{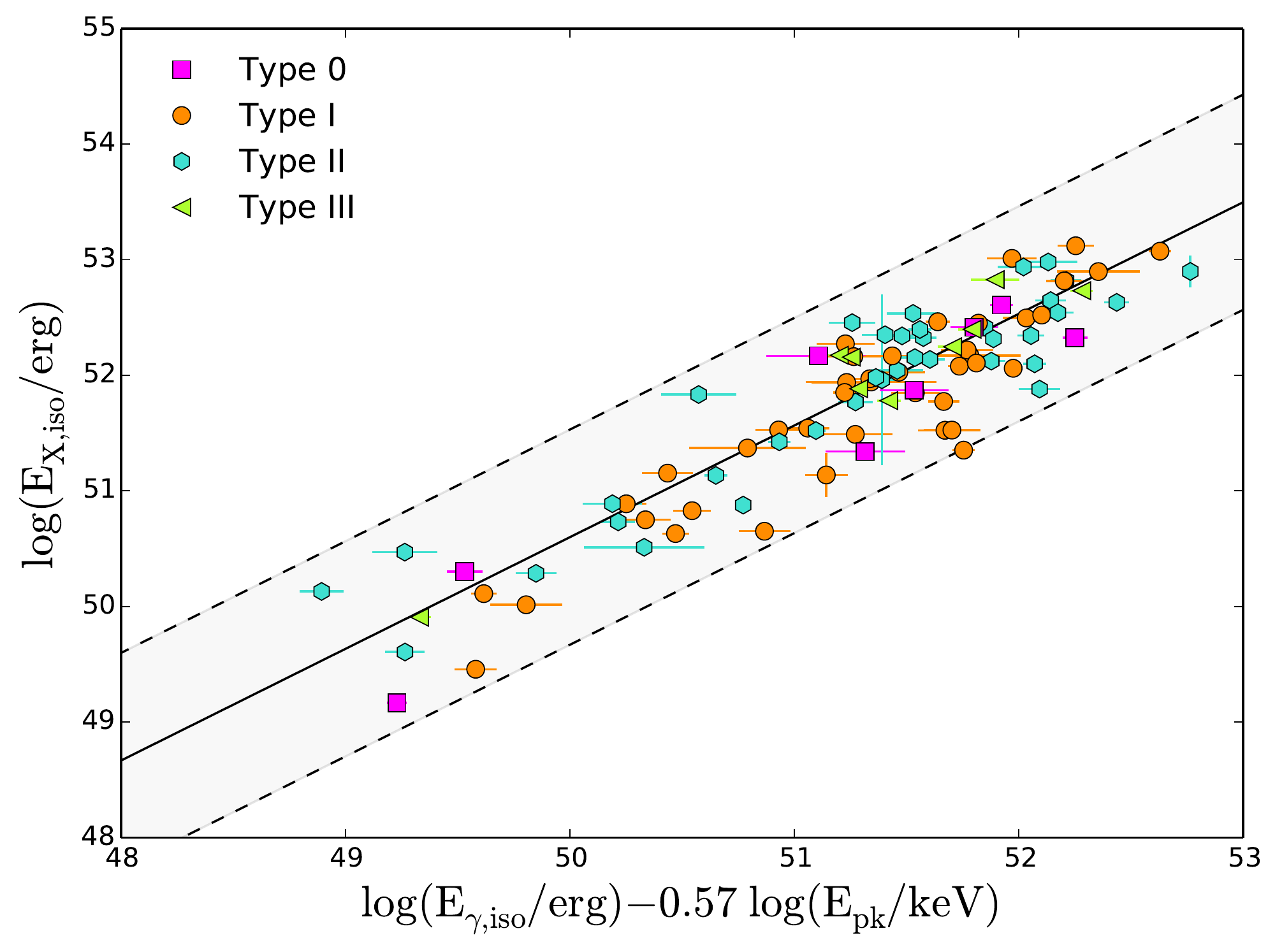}
\caption{E$_{\rm{X,iso}}$-E$_{\rm{\gamma,iso}}$-E$_{\rm{pk}}$ correlation, considering the different type of X-ray light-curves. \textit{pink squares}: Type 0; \textit{orange dots}: Type I; \textit{cyan hexagons}: Type II; \textit{green triangles}: Type III. The \textit{black solid line} is the best fit function, as in Figure \ref{fig:3par}, and the \textit{gray area} marks the 2$\sigma$ regions.}
\label{fig:type}
\end{figure}

We consider the possibility that the E$_{\rm{X,iso}}$-E$_{\rm{\gamma,iso}}$-E$_{\rm{pk}}$ correlation could be influenced by the different morphology of the X-ray light-curve. We classify the X-ray light-curves base on the number of break times in their fitting function (M13): \textit{Type 0} is fitted with a  single power-law, \textit{Type I} with the sum of two power-laws, \textit{Type II} (or \textit{canonical}) with the sum of three power-laws, and \textit{Type III} with the sum of four power-laws. Figure \ref{fig:type} shows the distribution of GRBs in E$_{\rm{\gamma,iso}}$ - E$_{pk}$ - E$_{\rm{X, iso}}$ plane depending on their X-ray light-curve morphology. Since GRBs with different X-ray light-curve shape are equally distributed on the plane, we conclude that the distribution of GRBs in the E$_{\rm{X,iso}}$-E$_{\rm{\gamma,iso}}$-E$_{\rm{pk}}$ plane is not dependent on this feature. In addition, we make the fit considering only the sub-group of canonical X-ray light-curves (33 GRBs) and we obtain fit parameters that are consistent with the values computed for the entire sample\footnote{For the subgroup of the canonical X-ray light-curves we calculate: $\rm{Log[E_{X,iso}}] =  (0.87\pm0.12) \rm{Log[E_{\gamma,iso}}] - (0.34\pm0.24) \rm{Log[E_{pk}]} -0.62 \pm 0.13$; if we fix $m_2=-0.59$ with $\sigma=0.28\pm0.05$.}. Therefore this correlation is not influenced by the different shapes of the X-ray light-curves.


\begin{table*}
\caption{From left to right: X and Y parameters to be correlated [the best-fitting law reads Log(Y) = q + m Log(X)]; best-fitting parameters as obtained accounting for the sample variance \citep{2005physics..11182D}: slope (m), normalization (q), intrinsic scatter ($\sigma$); errors are given at 95 per cent c.l. For each parameter couple, values reported in the first line refer to the entire sample, while in the second line we restrict our analysis to the long GRB class. E$_{\rm{X, iso}}$ and E$_{\rm{\gamma, iso}}$ are normalized to 10$^{52}$ erg, while E$_{\rm{pk}}$ to 100 keV.}
\label{tab:relations}
\begin{tabular}{c c c c c}
\hline
X & Y & m & q & $\sigma$\\
\hline
E$_{\rm{\gamma,iso}}$ & E$_{\rm{X,iso}}$ & 0.80$\pm$0.06 &  -0.84$\pm$0.08 & 0.40$\pm$0.04\\
                 &                & 0.68$\pm$0.06 &  -0.64$\pm$0.08 & 0.32$\pm$0.04\\

E$_{\rm{\gamma,iso}}$ & E$_{\rm{pk}}$   & 0.26$\pm$0.03 &   0.38$\pm$0.07 & 0.38$\pm$0.04\\
                  &               & 0.51$\pm$0.04 &  0.04$\pm$0.05  &0.19$\pm$0.02\\
                 
E$_{\rm{pk}}$       & E$_{\rm{X,iso}}$ & 0.57$\pm$0.25 &  -0.55$\pm$0.19 & 0.83$\pm$0.08\\
                  &               & 1.01$\pm$0.14 &  -0.53$\pm$0.10 & 0.40$\pm$0.05\\
 
E$_{\rm{pk}}$       & $\eta$        &-0.63$\pm$0.10 &  -0.62$\pm$0.08 & 0.32$\pm$0.04\\
                  &               &-0.57$\pm$0.11 &  -0.65$\pm$0.08 & 0.31$\pm$0.04\\
\hline
\end{tabular}
\end{table*}



\subsection{Physics and models}
\label{sec:physic}
As in B12 and M13, we defined $\epsilon$ = E$_{\rm{X,iso}}$/E$_{\rm{\gamma,iso}}$. $\epsilon$ represents the opposite of the efficency, that is  the ratio of the prompt emission energy and the outflow kinetic energy (e.g., \citealt{2004ApJ...613..477L}). As we showed in our previous works (B12, M13), we can divide the plane into two parts: one of the low-energetic GRBs which are less efficient and occupy the top left part of the plane, and the other group composed of  short and long GRBs, which have similar efficiencies. For this reason, from this plot it would be impossible to discriminate if GRB 090426A and GRB 100816A are long or short GRBs (\textit{gray squares} in Figure \ref{fig:efficiency}).

The physics behind the prompt emission is still one of the main open issues in the  study of GRBs, with several emission mechanisms and scenario being proposed. Among these, the  photospheric models (e.g. \citealt{2000ApJ...530..292M}) and the cannonball (CB) model (e.g. \citealt{2004PhR...405..203D}) are the ones providing most naturally a physical ground to the E$_{\rm{X,iso}}$-E$_{\rm{\gamma,iso}}$-E$_{\rm{pk}}$ correlation.

The photospheric model considers how the GRB spectrum in the optically thick phase can be modified by the interaction of the radiation field with the leptonic component of the outflow, before it is released at the
photosphere \citep{2000ApJ...530..292M, 2005ApJ...628..847R, 2006A&A...457..763G, 2009ApJ...700L..47L}. The simulations made by \citet{2013ApJ...765..103L} can reproduce the E$_{\rm{X,iso}}$-E$_{\rm{\gamma,iso}}$-E$_{\rm{pk}}$ correlation since the radiative efficiency of
brighter bursts is higher than that of weaker bursts. However, for adequately comparing observations and simulations, it is necessary to assume a value for the electron equipartition parameter $\epsilon$ \citep{2013ApJ...765..103L}.
They show that, by adopting the fiducial value $\epsilon$ = 0.1, a good agreement between simulation results and observed values is obtained.

In the CB model \citep{2004PhR...405..203D,2009ApJ...696..994D,2009ApJ...693..311D}, the Inverse Compton scattering caused by the interaction between the electrons of the CB plasma and the light in the near ambient of the SN is responsible of the $\gamma$-ray prompt emission of the GRBs, while the afterglow emission is related to the synchrotron radiation of the electrons swept-in and accelerated in the CBs. In this model, the E$_{\rm{X,iso}}$-E$_{\rm{\gamma,iso}}$-E$_{\rm{pk}}$ correlation is simply the combination of the two parameter correlations of kinetic origin that are followed by both long and short GRBs, even if with different normalizations, and so it depends on the large Doppler boosting and the relativistic beaming that strongly influenced the observed radiation \citep{2013ApJ...775...16D}. 


\section{Summary and conclusions}
\label{sec:summary}
In this paper we confirm the existence of the E$_{\rm{X,iso}}$-E$_{\rm{\gamma,iso}}$-E$_{\rm{pk}}$ correlation by employing a large sample of 94 GRBs  (35\% more than the previous sample and a double number of short GRBs, B12 and M13). The main feature of this correlation is that it involves both prompt and afterglow quantities (E$_{\rm{X,iso}}$, E$_{\rm{\gamma,iso}}$ and E$_{\rm{pk}}$) and it is followed by all kinds of GRBs, both short and long GRBs. 

As underlined in previous papers (B12, M13), this correlation implies the existence of common properties between long, short and low energetic GRBs, even if they have different progenitors and environments.

In particular, in this paper we have shown that:

\begin{enumerate}
\item The E$_{\rm{X,iso}}$-E$_{\rm{\gamma,iso}}$-E$_{\rm{pk}}$ correlation is followed by ultra-long GRBs (060218, 101225A, 111209A, and 130925A), which do not occupy a particular region in the plane. Indeed GRBs 101225A, 111209A, and 130925A behave as common long GRBs.
\item There is a possible intermediate group of transition between long and short GRBs, composed by different kinds of GRBs. In particular, in this group we find GRBs 090426, 100816A, and 060614 which have uncertain classification since they have properties of both long and short GRBs, and GRB 070714B that is a short GRB with EE.
\item We considered the possibility that the correlation could be biased by some assumption or evolve with some parameter. In B12 and M13 we excluded that the E$_{\rm{X,iso}}$-E$_{\rm{\gamma,iso}}$-E$_{\rm{pk}}$ correlation is influenced by the definition of E$_{\rm{X,iso}}$. Here, we showed that this relation is independent from the X-ray light curve morphology, indicating its robustness. A deeper analysis is needed to confirm its independence from the redshift.
\item As discussed in B12, the E$_{\rm{X,iso}}$-E$_{\rm{\gamma,iso}}$-E$_{\rm{pk}}$ correlation can be expressed  in the form of a two-parameter correlation between the GRB efficiency and E$_{\rm{pk}}$, as shown in Figure \ref{fig:efficiency}. The physical origin of such a relation may be connected with the outflow Lorentz factor.
\item The photospheric model  \citep{2013ApJ...765..103L} and the CB model \citep{2013ApJ...775...16D} can reproduce this correlation. 
\end{enumerate}

This updated sample confirms the existence of the E$_{\rm{X,iso}}$-E$_{\rm{\gamma,iso}}$-E$_{\rm{pk}}$ correlation. More data are necessary to confirm the possible existence of the intermediate group and to understand the possible physical processes that lead E$_{\rm{X,iso}}$, E$_{\rm{\gamma,iso}}$ and E$_{\rm{pk}}$ to be linked.

\section*{Acknowledgments}

We thank the anonymous referee for the helpful comments that have improved this paper. This work made use of data supplied by the UK Swift Science Data Centre at the University of Leicester. EZ thanks Luca Izzo and Marco Muccino for sharing their data. EZ acknowledges the support by the International Cooperation Program CAPES-ICRANet financed by CAPES - Brazilian Federal Agency for Support and Evaluation of Graduate Education within the Ministry of Education of Brazil. MGB thanks support from T-Rex Program.

\newcommand{\apj}{ApJ}
\newcommand{\mnras}{MNRAS}
\newcommand{\aap}{A\&A}
\newcommand{\araa}{ARA\&A}
\newcommand{\apjs}{ApJS}
\newcommand{\aj}{AJ}
\newcommand{\nat}{Nature}
\newcommand{\pasp}{PASP}
\newcommand{\apss}{Ap\&SS}
\newcommand{\aaps}{A\&AS}
\newcommand{\jgr}{J.Geophys.Res.}
\newcommand{\apjl}{ApJ}
\newcommand{\pra}{Phys.Rev.A}
\newcommand{\solphys}{Sol.Phys.}
\newcommand{\ssr}{Space Science Reviews}
\newcommand{\physrep}{Physics Reports}
\newcommand{\pasj}{PASJ}
\newcommand{\nar}{NewAR}
\bibliographystyle{mn2e} 
\bibliography{3par_v10} 

\appendix

\section{D'Agostini's method}
\label{agostini}
For the fit of the correlation that depends by three parameters, E$_{\rm{X,iso}}$, E$\rm{\gamma,iso}$ and E$_{\rm{pk}}$, we use the method of \citet{2005physics..11182D} (see Eq. (70) therein):
\begin{eqnarray}\nonumber
f(p) =&  0.5 \sum\log[p[4]^2 + \sigma(E_{\rm{pk}})^2 + (p[1] \sigma(E_{\rm{\gamma,iso}}))^2 + (p[2] \sigma(E_{\rm{X,iso}}))^2] + \\ \nonumber
          &  + 0.5 \sum{\frac{(E_{\rm{pk}} - p[1] E_{\rm{\gamma,iso}} - p[2]  E_{\rm{X,iso}} -p[3])^2 }{p[4]^2 + \sigma(E_{\rm{pk}})^2 + (p[1] \sigma(E_{\rm{\gamma,iso}}))^2 + (p[2]\sigma(E_{\rm{X,iso}}))^2}},\nonumber
\end{eqnarray}
with p[1], p[2] and p[3] the coefficients of the function and p[4] the extra-scatter parameter.

\begin{table*}
\caption{List of 33 GRBs added to the old sample. Short GRBs are marked in boldface, while the GRB with uncertain classification is underlined. a) Lower limit \citep{2014MNRAS.442.2342D}.}
\label{tab:sampletot}
\begin{tabular}{lcccc}
\hline
GRB & $z$ & E$_{\rm{X,iso}}$ (10$^{52}$) & E$_{\rm{\gamma,iso}}$ (10$^{52}$ erg) & E$_{\rm{pk}}$ (kev)\\
\hline
\textbf{080123}          & 0.495 & 0.0134$\pm$0.0002 & 0.13$^{a}$ & 149.50\\
\underline{090426}     & 2.609 & 0.142$\pm$0.011 & 0.541$\pm$0.064 & 320$\pm$54\\ 
\textbf{100117A}       & 0.920 &0.020$\pm$0.001 & 0.130$\pm$0.015 & 551$\pm$135\\
\textbf{100625A}       & 0.452 &0.0015$\pm$0.0002 & 0.075$\pm$0.003 & 701$\pm$114\\   
110106B                      & 0.618 & 0.007$\pm$0.003 & 0.734$\pm$0.073 & 194$\pm$56\\
110205A                      & 2.220 & 8.657$\pm$0.075 & 48.317$\pm$6.38 & 757$\pm$305\\
110213A                      & 1.460 & 2.243$\pm$0.042 & 5.778$\pm$0.813 & 224$\pm$74\\
110503A                      & 1.613 & 1.198$\pm$0.016 & 20.817$\pm$2.082 & 551$\pm$60\\
110715A                      & 0.820 & 0.765$\pm$0.016 & 4.361$\pm$0.445 & 220$\pm$22\\
110731A                      & 2.830 & 4.058$\pm$0.064 & 49.464$\pm$4.946& 1164$\pm$116\\
110801A                      & 1.858 & 0.742$\pm$0.021 & 10.897$\pm$2.724 & 400$\pm$171\\
110818A                      & 3.360 & 0.594$\pm$0.021 & 26.642$\pm$2.756 & 1116$\pm$240\\
111107A                      & 2.893 & 0.349$\pm$0.023 & 3.757$\pm$0.550 & 420$\pm$124\\
\textbf{111117A}       & 1.200 & 0.010$\pm$0.001 & 0.338$\pm$0.106 & 966$\pm$322\\      
111209A                      & 0.677 & 4.430$\pm$0.094 & 5.139$\pm$0.620 & 520$\pm$89\\
111228A                      & 0.716 & 0.603$\pm$0.008 & 2.750$\pm$0.275 & 58$\pm$7\\
120119A                      & 1.728 & 1.326 $\pm$0.030 & 27.197$\pm$3.626 & 496$\pm$50\\
120326A                      & 1.798 & 1.435$\pm$0.027 & 3.267$\pm$0.327 & 152$\pm$15\\
120712A                     & 4.174 & 1.764$\pm$0.126 & 21.199$\pm$2.110 & 641$\pm$130\\
120802A                     & 3.796 & 0.336$\pm$0.0150 & 12.886$\pm$2.761 & 274$\pm$93\\
120811C                     & 2.671 & 2.192$	\pm$0.061 & 6.405$\pm$0.640 & 198$\pm$20\\
121128A                     & 2.200 & 2.489$\pm$0.066 & 8.659$\pm$0.866 & 243$\pm$24\\
130408A                     & 3.758 & 1.276$\pm$0.126 & 34.972$\pm$6.442 & 1000$\pm$140\\
130427A                     & 0.340 & 3.480$\pm$0.0220 & 91.891$\pm$13.127 & 1250$\pm$150\\
130505A                     & 2.270 & 11.848$\pm$0.112 & 346.586$\pm$34.659 & 2030$\pm$203\\
\textbf{130603B}       & 0.356 & 0.0129$\pm$0.0005 & 0.212$\pm$0.023&  966$\pm$322\\
130701A                     & 1.155 & 0.332$\pm$0.008 & 0.415$\pm$0.041 & 2.283$\pm$0.020\\
130831A                     & 0.479 & 0.075$\pm$0.003 & 0.126$\pm$0.023 & 1.908$\pm$0.032\\
130907A                     & 1.238 & 7.921$\pm$2.480 & 2.465$\pm$0.027 & 2.944$\pm$0.013\\
130925A                     & 0.347 & 2.067$\pm$0.042 & 1.266$\pm$0.020 & 2.387$\pm$0.023\\
131030A                     & 1.295 & 1.150$\pm$0.020 & 1.482$\pm$0.026 & 2.609$\pm$0.023\\
140206A                     & 2.730 & 6.560$\pm$0.078 & 1.734$\pm$0.079 & 2.651$\pm$0.035\\
140419A                     & 3.956 & 7.896$\pm$0.092 & 2.180$\pm$0.171 & 3.160$\pm$0.109\\
\hline
\end{tabular}
\end{table*}

\end{document}